\documentclass[10pt,a4paper,showpacs,english,aps,prl,twocolumn]{revtex4}
\usepackage[OT1]{fontenc}
\usepackage[latin1]{inputenc}
\usepackage{amsmath}
\usepackage{graphicx}
\usepackage{amssymb}

\makeatletter

\renewcommand{\vec}[1]{\mathbf{#1}}
\renewcommand{\Re}[1]{\mathrm{Re\,}}
\renewcommand{\Im}[1]{\mathrm{Im\,}}

\newcommand{\up}[1][]{_{\uparrow #1}}
\newcommand{\down}[1][]{_{\downarrow #1}}
\newcommand{\LD}{\epsilon}
\newcommand{\SP}{\xi}

\usepackage{babel}
\begin{document}

\title{Finite temperature phase diagram of a polarized Fermi gas in an optical lattice}
\author{T.K. Koponen$^1$, T. Paananen$^2$ , J.-P. Martikainen$^2$ , and P. T\"{o}rm\"{a}$^1$}
%\email{timo.koponen@phys.jyu.fi}
\email{paivi.torma@phys.jyu.fi}
\affiliation{$^1$Department of Physics, Nanoscience Center, P.O.Box 35, 40014
University of Jyv\"{a}skyl\"{a}, Finland\\$^2$ Department of Physical Sciences, 
P.O.Box 64, 00014 University of Helsinki,  Finland}

%Huom:74.25.Dw=Phase diagrams,superconductors
%05.30.Jp=Boson systems joten sitä ei kannata käyttää

\begin{abstract}
We present phase diagrams for a polarized Fermi gas in an optical
lattice as a function of temperature, polarization, and lattice filling
factor. We consider the Fulde-Ferrel-Larkin-Ovchinnikov (FFLO), Sarma or
breached pair (BP), and BCS phases, and the normal state
and phase separation. We show that the FFLO phase appears in a
considerable portion of the phase diagram. The diagrams have two
critical points of different nature. We show how various phases leave
clear signatures to momentum distributions of the atoms which can be
observed after time of flight expansion.
\end{abstract}

\maketitle

Recent advances in the experiments of ultracold Fermi gases have shown great potential for 
elucidating long-standing problems in many different fields of physics related to strongly correlated %%@
Fermions. For instance, 
in recent
experiments~\cite{Zwierlein2006a,Partridge2006a,Zwierlein2006c,Shin2006a,Partridge2006c} 
spin-density imbalanced, or polarized, Fermi gases were
considered. %Among other things, s
Such systems %%
make it
possible to study pairing with mismatched Fermi surfaces, potentially
leading to non-standard phases 
such 
as that appearing in FFLO-states~\cite{Fulde1964a,Larkin1964a} or BP-states~\cite{Sarma1963a} %%@
(Sarma-states). These possibilities have been considered 
extensively in condensed-matter, nuclear, and high-energy physics~\cite{Casalbuoni2004a}. 
The experiments in trapped gases have shown clear evidence of the separation 
of the gas into a BCS core region and a normal state shell around it, i.e.~phase separation. %%@
Although an FFLO-type state has 
been predicted to appear in these systems as
well~\cite{Mizushima2005b,Machida2006a}, 
it is likely to be difficult to observe %%@
since it appears in the edges
of the trap except for large polarizations. 

\begin{figure}
\includegraphics[width=0.84\columnwidth]{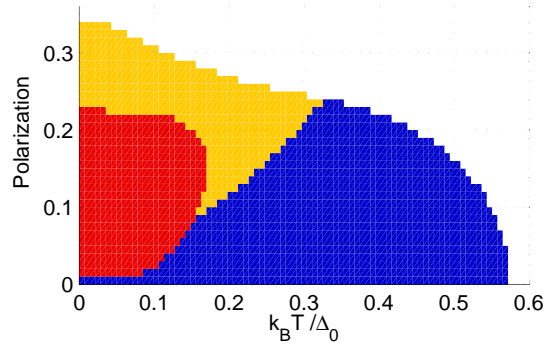}
\caption{(Color online) The phase diagram of the imbalanced Fermi gas in a
lattice. %as a function of temperature and polarization. 
Colors: 
BP $=$ blue (at $P=0$ this is actually BCS), FFLO $=$ yellow, PS $=$ red, normal $=$ white. 
The average filling %factor was  
is $0.2$ atoms/lattice site in
each component, $J=0.07E_R$, and $U=-0.26 E_R$, where $E_R=\hbar^2k^2/2m$ is the
recoil energy, $k=2\pi/\lambda$ and %the laser wavelength was
$\lambda=1030\, {\rm nm}$. Here $\Delta_0$ means the gap at $T=0$ and
$P=0$. The $T_{\text{C}}$ at $P=0$ is $41$ nK.}
\label{Fig1}
\end{figure}

%Fermi gases confined in optical %%@ lattices~
Recent experiments~\cite{Kohl2006a,Chin2006a} on Fermi gases confined
in optical lattices have already demonstrated the potential of these
systems for a multitude of studies of new phases, dimensionality
effects, and dynamics. % are expected to offer 
%a multitude of new possibilities for studies of new phases,
%dimensionality effects and dynamics. 
In this letter, we calculate the 
phase diagram for an attractively interacting Fermi gas in an optical lattice at zero and finite %%@
temperatures. 
%In particular, 
%we consider the possibility of the single mode FFLO-phase and BCS-BP-phase as well as their competition %%@
%with the phase separation (PS) of 
%the gas into normal and BCS superfluid regions.  
In particular, we consider the possibility of the single mode
FFLO-phase, where the order parameter is space-dependent, and
of the BCS-BP phase, where compared to the standard BCS phase
the excess polarization is carried by additional Bogoliubov
excitations. We investigate their competition with the phase
separation (PS) of the gas into normal and BCS superfluid regions.
Our results reveal that a typical 
phase diagram as a function of polarization and temperature is as shown in Fig.~\ref{Fig1}: phase %%@
separation is expected for small polarizations
and temperatures, whereas at zero temperature the FFLO state appears after some critical %%@
polarization. At a finite temperature, the FFLO-phase competes 
with the BCS-BP phase. As the temperature increases the BCS-BP phase becomes energetically favorable, and as %%@
the temperature
is increased even further the BCS-BP phase gives way to a normal polarized
Fermi gas. 

Furthermore, the phase diagrams reveal 
a Lifshitz point which is
surrounded by the normal, FFLO, and BCS-BP phases. The transitions around
this point are of second order, but the FFLO phase breaks the
translational symmetry which is present in BCS-BP and
normal phases. Interestingly, there is also a point surrounded
by the PS domain and the FFLO and BCS-BP phases. Comparing this point to the tricritical
point discussed in Refs.~\cite{Gubbels2006b,Parish2007a}, the difference is
that the part of the phase diagram around the tricritical point which
was occupied by the normal state in Refs.~\cite{Gubbels2006b,Parish2007a}, is now
occupied by the symmetry breaking FFLO phase. If extending to stronger
interactions, the order of the transitions in question should be
considered more carefully.

In atomic gases, the mismatch of the Fermi surfaces is realized by a fixed atom number difference %%@
between 
the two components, i.e. fixed polarization. This 
is in contrast to many other physical systems (such as
superconductors or quark matter) where it is caused by different chemical potentials, due to e.g.
an external magnetic field. In such systems it is known that, at zero temperature, the BCS state is %%@
realized for small chemical 
potential differences, and after a critical difference, some exotic pairing state such as FFLO or BP %%@
state may appear, although the 
stability of these states is under a debate~\cite{Bedaque2003a,Gubankova2004a,Sheehy2006a}. If the polarization, %%@
rather than the chemical potential difference, is fixed, 
the BCS state cannot be the solution at zero temperature. Consequently, it is important 
to consider the phase separated
state as an alternative to the FFLO and BP states. In fact, it turns out that at zero temperature the %%@
phase separation into
%a phase with 
an unpolarized BCS region and a polarized normal region gives a lower energy than the BP %%@
phase~\cite{Bedaque2003a}.

The phase diagrams including the possibility of FFLO-phase and phase separation have %%@
been calculated for atomic Fermi 
gases in free space~\cite{Sheehy2006a,He2006b}. Also there, like in our results, the phase separation appears for low %%@
polarizations and the FFLO phase for higher ones. However, a striking difference is that our calculations in optical 
lattices give a considerable parameter window for the existence 
of the FFLO phase, whereas in free space the parameter window is extremely narrow. 
Recently, density polarized gases in optical lattices were
considered in Ref.~\cite{Iskin2006b} but that work concentrates 
on the insulating phases and does not resolve the competition between FFLO, BCS-BP, and phase separation, %%@
moreover the discussion there is limited to zero temperature. 
Finite temperature phase diagrams for trapped imbalanced gases have
been 
considered in~\cite{Machida2006a,Martikainen2006a,Gubbels2006b,Chien2007c}.
These works do not consider the FFLO state, except~\cite{Machida2006a}. There a FFLO-type phase
appears via the Bogoliubov-de Gennes ansatz, not via a plane wave ansatz customary in  
translationally invariant or periodic systems, therefore the spatial behaviour of the order
parameter is different: for small and intermediate polarizations the oscillations of the
order parameter appear only in the edges of the trap.

We consider two distinguishable species of fermionic atoms, e.g. atoms in different hyperfine states, 
labelled by up and down pseudospin index, 
interacting via an attractive contact interaction and confined in an optical lattice. The system is 
described by the Hamiltonian 
\(H =\sum_{\vec{k}}(\sum_{\sigma}(\LD_{\vec{k}} - \mu_\sigma )\hat{c}_{\sigma,\vec{k}}^\dagger 
\hat{c}_{\sigma,\vec{k}} +
\Delta \hat{c}\up[\vec{q}+\vec{k}]^\dagger
\hat{c}\down[\vec{q}-\vec{k}]^\dagger + h.c.)
 -\frac{|\Delta|^2}{U}\), where \(\LD_{\vec{k}} =
 2\sum_{i=1}^{3}J_i(1-\cos(k_i d))\). Here $\vec{q}$ is the FFLO
 momentum and it represents the momentum of the Cooper pairs. In the single
 mode
ansatz the spatial dependence of the order parameter is given by
$\Delta({\bf r})=\Delta \exp(2i{\bf q}\cdot {\bf r})$ .
The parameters $U$ and $J_i$ are defined as in
\cite{Jaksch1998a}. Our calculations are in the intermediate and weak
coupling regime ($U/J < 6$), where the single band approximation
is valid.
With $\vec{q} = 0$ the Hamiltonian is the standard BCS-BP
 Hamiltonian. The quasiparticle energies of the system are \( E_{\vec{k},\vec{q},\pm} = %%@
\frac{\SP\up[\vec{k}+\vec{q}] -
    \SP\down[-\vec{k}+\vec{q}]}{2} \pm \sqrt{ \left(\frac{\SP\up[\vec{k}+\vec{q}] +
    \SP\down[-\vec{k}+\vec{q}]}{2}\right)^2 + \Delta^2}\), where
\(\SP_{\sigma,\vec{k}} = \LD_{\vec{k}} - \mu_\sigma\). We consider
fixed numbers of particles and determine the chemical potentials by
the number equations \(N\up =  \sum_{\vec{k}}
  u_{\vec{k},\vec{q}}^2 f(E_{\vec{k},\vec{q},+}) + v_{\vec{k},\vec{q}}^2
  f(E_{\vec{k},\vec{q},-})\) and \(N\down = \sum_{\vec{k}}
  u_{\vec{k},\vec{q}}^2 f(-E_{\vec{k},\vec{q},-}) + v_{\vec{k},\vec{q}}^2
  f(-E_{\vec{k},\vec{q},+})\), where \(f\) is the Fermi function and the coefficients $u$
and $v$ are defined in the usual way, see for example
\cite{Koponen2006b}. The
  grand potential of the system is \(\Omega = \sum_{\vec{k}}\left[ \SP\down[-\vec{k}+\vec{q}] +
  E_{\vec{k},\vec{q},-} -
    \frac{\Delta^2}{U}\right] - \frac{1}{\beta} \sum_{\vec{k}}\left[\ln\left(1 + e^{-\beta
      E_{\vec{k},\vec{q},+}} \right) + \ln\left(1 + e^{\beta
      E_{\vec{k},\vec{q},-}} \right)\right]\). For any given point in
the phase diagrams, we look for the stable phase by minimizing the
free energy \(F = \Omega + \mu\up n\up + \mu\down n\down\) with respect to $\vec{q}$ and $\Delta$, %%@
where
$n\up$ and $n\down$ are the filling factors of the different atoms. We
define the different phases as follows: the region of the phase diagram where the energy is
minimized by $\Delta = 0$ is normal phase, the region where $\Delta >
0$ and $q > 0$ is FFLO. The regions where $\Delta > 0$ and $q = 0$ are
BP, and within the BP region, the line with no polarization, i.e. $n\up
= n\down$, is the standard BCS state. All of these phases are obtained as
stable solutions from the free energy given above and the BCS state 
can be reached smoothly from the BP state by reducing the polarization
to zero.

We also compare all the above states
to a phase separation between the standard BCS state and the polarized
normal state. We do this as in Ref.~\cite{Bedaque2003a}:
with numbers $N\up$ and $N\down$ in the different components, we let
$N_{BCS}$ atoms from both components occupy the BCS state and put the rest
of the atoms to the normal state. We let the normal state occupy a
fraction $x$ of the lattice, leaving the fraction $1-x$
for the BCS state. We then add the free energies of a BCS state with a
density $n_{BCS}/(1-x)$ in both components and a normal
gas with densities $(n\up - n_{BCS})/x$ and $(n\down - n_{BCS})/x$ in the
different components. Thereafter we vary both $N_{BCS}$ and $x$ to find the
energy of the optimal phase separated state and compare it to the
optimal free energy given by a single state.

\begin{figure}
\includegraphics[width=0.84\columnwidth]{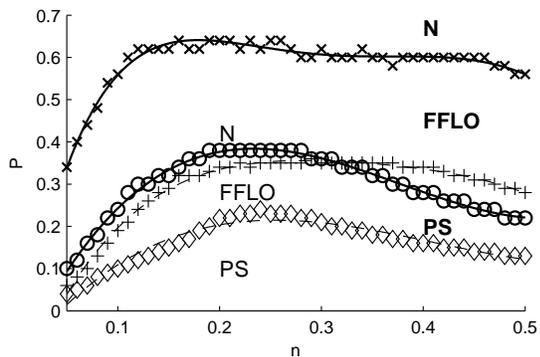}
\caption{Zero temperature phase diagram as a function
of average filling $n$ and polarization $P$ for two different interaction
strengths. The transition lines are with $U=-0.36 E_R$ (solid) and
$U=-0.26 E_R$ (dashed). We indicate the actual data points and the
lines are meant as guides to the eye. We choose  $J=0.07E_R$, $E_R$ is the same as before.}

%The solid lines give the transition lines 
%with $U=-0.36 E_R$ while dashed lines give transitions 
%with  $U=-0.26 E_R$. We indicate the actual %numerical 
%data points
%and the lines are meant %should be intepreted 
%as guides to the eye. $J$ is $0.07E_R$.}
%We choose  $J=0.07E_R$, $E_R$ is the same as before.}
\label{Fig2}
\end{figure}

Fig.~\ref{Fig2} shows the zero temperature phase diagram as a function of polarization $P$ and the %%@
average lattice filling $n$, 
for two different interaction strengths. 
Starting from zero, the critical polarization for the FFLO-normal boundary increases as function of %%@
the average filling. The maximum 
critical polarization is achieved for a certain filling factor that is
the smaller the stronger %%@
the interaction strength is, and is different
from half filling. Note that although half filling is predicted to
give the maximum gap~\cite{Micnas1990a}, in our case
the value $0.5$ of the average filling $n$, for high polarizations, is
far from the actual half filling. In fact, for large $n$ and $P$ 
we would enter a regime outside the scope of our Hamiltonian
restricted to the lowest band.\footnote{A similar diagram with the normal phase boundary, but not 
the boundary between PS and FFLO, is given in~\cite{Iskin2006b}.} 
The phase separation - FFLO boundary 
shows interesting behaviour: both critical %%@
polarizations (for the PS-FFLO and
FFLO-normal) are close to each other for small fillings, then grow when the density increases, 
however, the PS-FFLO critical polarization starts %%@
to decrease earlier and stronger
than the FFLO-normal critical polarization. 
This implies that, for high densities, the polarization window for the FFLO phase can become
large compared to the polarization window for the PS domain. For small
densities the critical polarization tends to zero and the window
for FFLO phase becomes narrow, as expected in the long wavelength limit
where the lattice dispersion can be approximated with the free space
dispersion.

The stability of the FFLO state in a lattice is due to the flatness and
nesting of the majority and minority component Fermi surfaces. In the
continuum limit the Fermi surfaces are spheres, and the FFLO ansatz
effectively shifts them relative to each other, allowing matching of
parts of the Fermi surfaces. For small lattice fillings $n$ the well
known result 
$q\sim k_{\text{F}\uparrow}-k_{\text{F}\downarrow}$ gives a good
estimate for the $q$ we find numerically. For higher fillings the
surfaces deform from spherical towards octahedra. By shifting two
octahedra relative to each other, four whole faces of the minor
component octahedron can be made to match the major component Fermi
surface. Such nesting makes the stability of FFLO intuitively
understandable. For fillings close to $0.2$ where the Fermi surfaces
can be approximated by octahedra, we find that the numerically
obtained $q$ can be well estimated by the shift needed along the
diagonal of the octahedra required to make the surfaces
match. For intermediate cases between spherical and octahedron, we
made numerical optimization of the match of the corresponding ideal
gas Fermi surfaces and found again good agreement with the values of
$q$ given by the full FFLO calculation. Typical values given for
$qd$ by our FFLO calculations are between $0$ and $0.4$, larger
fillings giving larger values of $q$.

Fig.~\ref{Fig1}  shows the finite temperature phase diagram. %As
                                %discussed earlier, t
The FFLO state 
dominates over the phase separation after some finite critical polarization is reached, even at zero %%@
temperature. At high enough 
temperatures the phase separation gives way to  the FFLO, BCS-BP, or
normal phases.  Note that the FFLO regime, %%@
both at zero and finite temperature,
is reasonably large, not vanishingly small as in many other systems, which makes %%@
the observation of this phase in optical lattices 
more feasible. We have calculated finite temperature phase diagrams for several values of the
interaction strength $U$ and the average density $n$. In general, larger values of these parameters produce
higher critical polarizations and critical temperatures for the FFLO and BP states. The PS-FFLO boundary has
a non-trivial dependence on the density, as already indicated by Figure 2. Overall, the qualitative 
behaviour presented in Figure 1 is not sensitive for the variation of
the parameters. For the parameters of Figure 1, $\Delta(P=0,T=0) \sim 0.17 E_F$ and
$T_c(P=0) \sim 0.1 E_F$. Including pair fluctuations \cite{Nozieres1985a} would shift the normal state boundary down
in temperature at higher temperatures but would not change the
qualitative features such as the large FFLO window found in lower
temperatures. We have checked that including Hartree corrections
does not change this result either. At $T=0$, the polarizations $P$
for which phase separation (PS) and FFLO appear are, for three
different interaction strengths $U/(6J) = U/W$, the following: phase
separation dominates for $P\sim [0:0.23]$ ($U/W = 0.61$, gap ($ =
\Delta(P=0,T=0)$) $\sim 0.17$~$E_\text{F}$, $P \sim [0:0.08]$ ($U/W =
0.42$, gap $\sim 0.06$~$E_\text{F}$), $P \sim [0:0.05]$ ($U/W = 0.37$,
gap $\sim 0.04$~$E_\text{F}$), and FFLO for $P \sim [0.23:0.34]$,
$[0.08:0.13]$, $[0.05:0.08]$, respectively, from where the relative
size of the FFLO domain versus the total PS+FFLO domain is obtained to
be $0.32$, $0.38$, $0.38$, respectively, i.e. it is essentially the
same also for smaller interactions (even slightly grows).

\begin{figure}
\includegraphics[width=0.41\columnwidth]{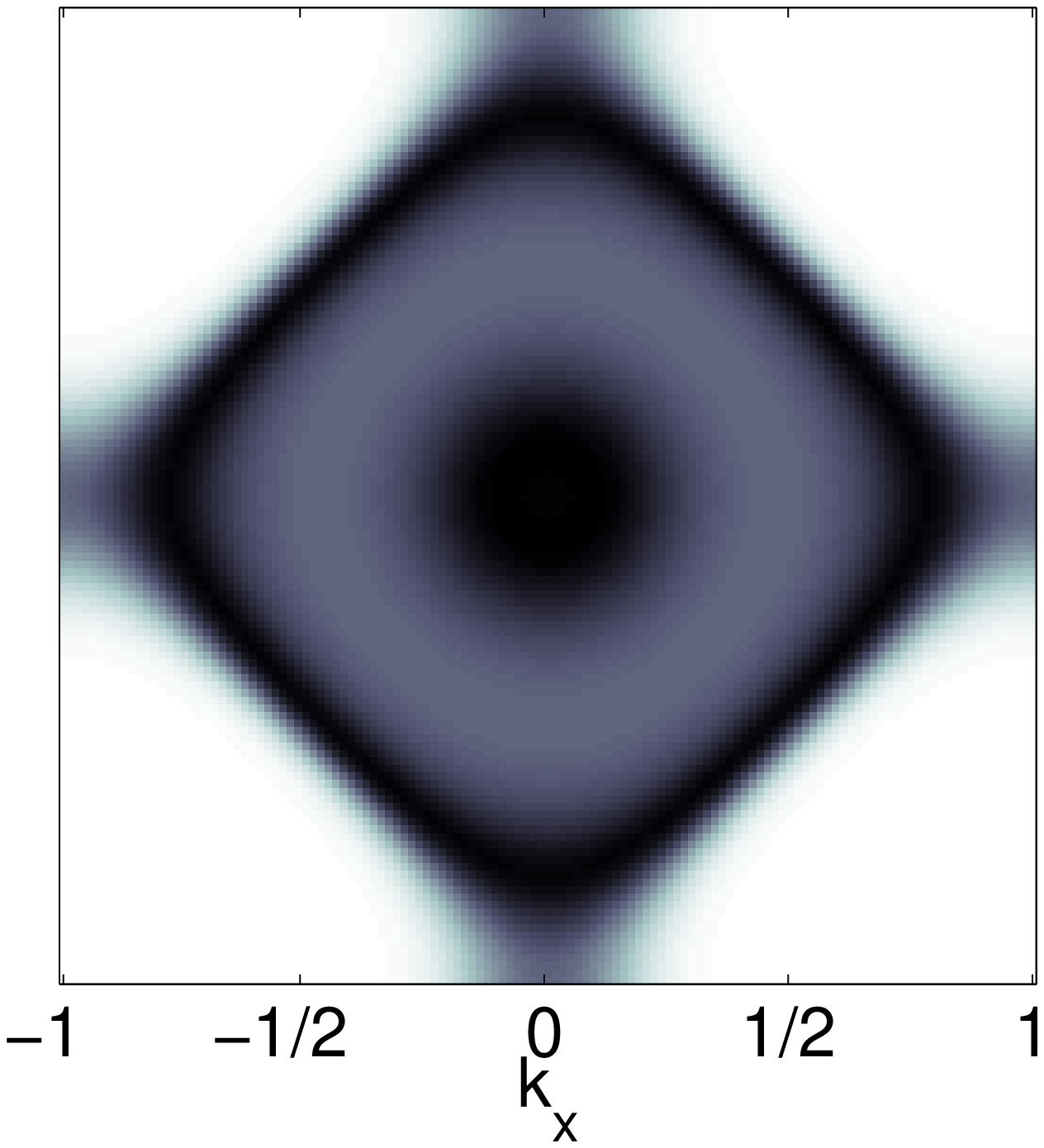}
\includegraphics[width=0.41\columnwidth]{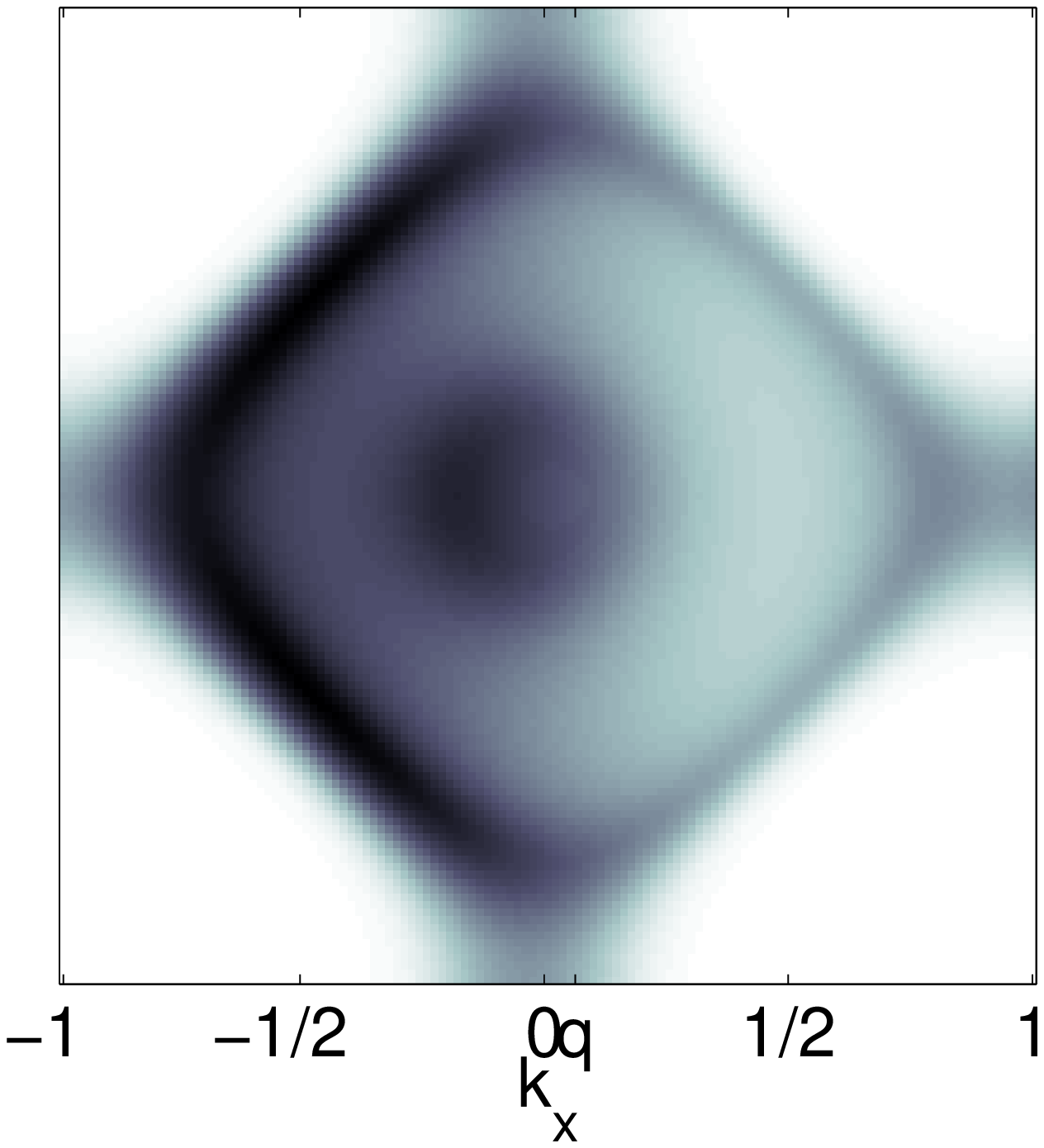}
\caption{(Color online) Difference in the momentum distributions,
  $n\up - n\down$, integrated over the z-direction, of the two
  components %species of fermions
  with an average filling of $0.2$
  atoms/site in the BP (left) and FFLO (right) states. Darker
  colour means larger difference. The $q$ is marked on the $k_x$ axis
  in the FFLO case.}
\label{Fig3}
\end{figure}

All the phases that we discussed here are directly observable in optical lattices from the momentum %%@
distributions that are obtained
by imaging after time of flight expansion.  Also other methods, such as observing
the noise correlations~\cite{Altman2004a}%,Greiner2005a,Foelling2005a} 
can be considered. 
Fig.~\ref{Fig3} shows how 
the FFLO (at $15$ nK) and BP (at $20$ nK) phases are reflected in the momentum %%@
distribution. A normal state would 
be completely symmetric in the $k$-space, in contrast to the FFLO, and the phase separated state can be assumed to produce a %%@
symmetric result as well (with a 
background trap, the actual form of phase separation is likely to be a BCS state in the middle %%@
of the trap and the normal gas 
on the edges). Furthermore, the phase separated state can be distinguished,
for example, by removing the paired atoms from the system using
RF-fields (detuning matched to the pairing gap to select the paired atoms only) and 
then observing the remaining cloud of atoms which is, after the removal,
expected to have a non-monotonous density distribution if
the initial state was a phase separated state.

The background trapping potential that is present in most optical lattice experiments would further
decrease the energy of the phase separated solution because of the higher density in the middle of %%@
the trap. The background trap must therefore be sufficiently slowly varying for the FFLO state to %%@
remain energetically favourable. An order of magnitude estimate for the maximum trapping frequency %%@
can be made by considering the energy difference between the FFLO and phase separation solutions, per %%@
particle, $\delta E$ which for typical parameters used here is about
$10^{-3} E_R$. We compare this to %%@
the potential energy of a harmonic trap $m\omega^2 x^2/2$ at the distance $x$ given by the size of %%@
the system, i.e. $x=N d/2$, where $N$ is the number of sites in one dimension and $d$ is the lattice %%@
spacing. This gives the maximum energy that is possible to save by moving a particle from the edge to the %%@
middle; we also multiply this by the polarization to correspond to the particles that would likely be %%@
affected by the phase separation. Comparing these two energies, for the parameters used here ($^6Li$, 
$d=0.5 \mu m$, $N=64$), %%@
we obtain $\omega_{max}$ of the order of $100 {\rm Hz}$, which is not unreasonable.  

In summary, we presented for the first time a finite temperature phase diagram for optical lattices %%@
where
various forms of superfluidity and phase separation are compared. Apart from giving the %relevant %%@
scales for critical polarizations and 
temperatures, relevant for the experiments, we find a clear qualitative difference to similar works %%@
in other systems: the parameter window
for the existence of FFLO can be rather large, the phase is certainly not
negligible or an edge effect. 
It turns out that the FFLO phase often appears in the parameter region where the lattice problem
cannot be approximated by the long wavelength limit, which explains the 
pronounced difference with the free space results, where the FFLO phase is
expected only in a very narrow parameter region.
Another remarkable difference between optical lattices 
and other systems realizing imbalanced Fermi systems is 
that the signatures of the different phases should be readily
observable. %Intriguingly, we also find a Lifshitz point surrounded
%by BCS-BP, FFLO and normal phases, as well as another critical point
%surrounded by the PS domain and the FFLO and BP phases.
We show that the FFLO and BCS-BP states are clearly reflected
in the momentum distribution~\cite{Koponen2006b}  of the gas, which can be directly observed after a %%@
time of flight expansion. %when the gas is released from the optical 
%lattice.
Our study has been restricted to a single mode FFLO ansatz and it is likely that
multi-${\bf q}$ FFLO superfluids have a lower energy. Therefore,
regions of our phase diagrams where FFLO is expected to appear
may be larger and also contain several different types
of FFLO phases. 

{\it Acknowledgements}
We thank J.~Kinnunen and L.M. Jensen for discussions.
This work was supported by the National Graduate School in Materials Physics and 
QUDEDIS and Academy of Finland (project numbers 106299, 115020,
213362, 207083)
and conducted as part of a EURYI scheme award. See www.esf.org/euryi.

%\paragraph*{}
\bibliographystyle{apsrev}

%\bibliography{paperi}
\end{document}